\documentclass[conference]{IEEEtran}
\IEEEoverridecommandlockouts
\usepackage{cite}
\usepackage{subcaption}
\usepackage{textcomp}
\usepackage{amsmath,amssymb,amsfonts}
\usepackage[ruled,vlined,linesnumbered]{algorithm2e}
\usepackage{mathtools}
\usepackage{graphicx}
\usepackage{xcolor}
\usepackage{fancyvrb}

\usepackage{amssymb}
\usepackage{lscape}
\usepackage{pdflscape}

\usepackage{color}
\usepackage{hyperref}
\usepackage{float} 
\usepackage{dirtytalk}
\usepackage{soul}
\usepackage[compact]{titlesec}

\PassOptionsToPackage{bookmarks=false}{hyperref}
\hypersetup{
    colorlinks=true,
    linkcolor=blue,
    filecolor=blue,      
    urlcolor=blue,
    citecolor=blue
}

\newtheorem{definition}{Definition}

\def\BibTeX{{\rm B\kern-.05em{\sc i\kern-.025em b}\kern-.08em
    T\kern-.1667em\lower.7ex\hbox{E}\kern-.125emX}}

\definecolor{darkgreen}{rgb}{0,0.5,0}

\definecolor{grigiomoltochiaro}{gray}{0.97}
\definecolor{verde}{rgb}{0,1,0}

\newcommand{\define}{\triangleq}

\begin{document}

\makeatletter
 \let\old@ps@headings\ps@headings
 \let\old@ps@IEEEtitlepagestyle\ps@IEEEtitlepagestyle
 \def\confheader#1{

 \def\ps@IEEEtitlepagestyle{
 \old@ps@IEEEtitlepagestyle
 \def\@oddhead{\strut\hfill#1\hfill\strut}
 \def\@evenhead{\strut\hfill#1\hfill\strut}
 }
 \ps@headings
 }
 \makeatother

\confheader{
Accepted at the 19th IEEE Intl. Conference on Trust, Security and Privacy in Computing and Communications (TrustCom 2020)
 }
\title{BONIK: A Blockchain Empowered Chatbot for Financial Transactions\\
}

\author{\IEEEauthorblockN{Md. Saiful Islam Bhuiyan\IEEEauthorrefmark{1}, Abdur Razzak\IEEEauthorrefmark{1}, Md Sadek Ferdous\IEEEauthorrefmark{1} \IEEEauthorrefmark{2}, Mohammad Jabed M. Chowdhury\IEEEauthorrefmark{3},\\Mohammad A. Hoque\IEEEauthorrefmark{4}, Sasu Tarkoma\IEEEauthorrefmark{4}}
\IEEEauthorblockA{\IEEEauthorrefmark{1}Department of Computer Science and Engineering, 
Shahjalal University of Science and Technology, Sylhet, Bangladesh\\}
\IEEEauthorblockA{\IEEEauthorrefmark{2}Imperial College Business School, Imperial College London, London, United Kingdom\\}
\IEEEauthorblockA{\IEEEauthorrefmark{3}Department of Computer Science \& Information Technology, La Trobe University, Melbourne, Australia\\}
\IEEEauthorblockA{\IEEEauthorrefmark{4}Department of Computer Science, University of Helsinki, Helsinki, Finland\\
Email: saif\_lesnar@outlook.com, razzakrana17@gmail.com, sadek-cse@sust.edu, m.chowdhury@latrobe.edu.au,\\\{mohammad.a.hoque, sasu.tarkoma\}@helsinki.fi\\}}

\maketitle
 
\begin{abstract}
A Chatbot is a popular platform to enable users to interact with a software or website to gather information or execute actions in an automated fashion. In recent years, chatbots are being used for executing financial transactions, however, there are a number of security issues, such as secure authentication, data integrity, system availability and transparency, that must be carefully handled for their wide-scale adoption. Recently, the  blockchain technology, with a number of security advantages, has emerged as one of the foundational technologies with the potential to disrupt a number of application domains, particularly in the financial sector. In this paper, we forward the idea of integrating a chatbot with blockchain technology in the view to improve the security issues in financial chatbots. More specifically, we present \textit{BONIK}, a blockchain empowered chatbot for financial transactions, and discuss its architecture and design choices. Furthermore, we explore the developed Proof-of-Concept (PoC), evaluate its performance, analyse how different security and privacy issues are mitigated using BONIK.
\end{abstract}

\begin{IEEEkeywords}
Blockchain, Chatbot, Financial Chatbot, Financial Transaction, Private Blockchain, Hyperledger Fabric
\end{IEEEkeywords}

\section{Introduction}
\label{chap:intro}
A chatbot is an advanced application of Artificial Intelligence (AI), providing users with a platform to interact with software or web services in an automated fashion. In recent times, the chatbots have emerged as a technology with a wide-scale adoption in the industry, particularly as an alternative for customer care services that require trivial interactions \cite{chatbot1}. A recent study finds that $74\%$ of users prefer to engage with chatbots for answers to simple queries \cite{chatbot2}. The increased adoption of chatbots has resulted in an ever-increasing market size, estimated to be $2.6B$ USD in 2019 with a forecast to increase to $9.4B$ USD by 2024 \cite{chatbotMSize}. Following this trend, the service industry is exploring the possibility of financial chatbots to facilitate financial transactions in a seamless fashion \cite{okuda2018ai}. For example, WeChat, a prominent Chinese message app with a chatbot facility, introduced a fund transfer facility in their messaging platform \cite{wechat}. However, transactions are much more sensitive than services answering to mere trivial queries. Therefore, such chatbots must guarantee the security and privacy properties of financial transactions \cite{bozic2018security, lai2018banking, yan2018identifying}, such as confidentiality, integrity, authenticity, availability, control, and transparency. Also, relying on a single entity to transfer funds introduces a single point of failure. These issues must be addressed before the wide-scale adoption of such chatbots.

In recent years, Blockchain technology (blockchain, in short) has emerged as one of the fundamental technologies with the potential to disrupt several application domains \cite{chowdhury2019comparative}. Blockchain offers several advantages such as immutability of data and code, distributed consensus mechanism, data provenance, and transparency \cite{ferdous2020blockchain}, which can effectively tackle the above-mentioned security issues for financial chatbots. Even though a few existing works (e.g., in \cite{bozic2018security, lai2018banking, yan2018identifying}) explored different security and privacy issues in chatbots, including WeChat, an effective solution is still at large. In this paper, we present \textit{BONIK}, 
a blockchain empowered chatbot for financial transactions that effectively addresses the security issues involving a financial chatbot. Using BONIK, one can execute financial transactions in a secure and privacy-friendly way by interacting with a chatbot. In this paper, we present its architecture, protocol flow, usages, and different other aspects.

\vspace{2mm}
\noindent \textbf{Contributions:} The main contributions of the paper are presented below:
\begin{itemize}
    \item We formulate several functional, security and privacy requirements, underpinned by a rigorous threat model, for a financial chatbot.
    \item We provide a detailed architecture of BONIK and discuss how we have developed a Proof-of-Concept (PoC) prototype along with its detailed protocol flow. 
    \item We evaluate BONIK's performance and analyse how the developed prototype satisfies the formulated requirements and explores its advantages and limitations.
\end{itemize}

\vspace{2mm}
\noindent \textbf{Structure:} Section \ref{sec:back} provides a brief background on blockchain and chatbot. Section \ref{sec:proposal} presents a threat model and requirement analysis. Section \ref{sec:archi} outlines the architecture of BONIK with implementation details. In Section \ref{sec:protocol}, the protocol flow of BONIK illustrates its use-case. Section \ref{sec:evaluation} evaluates the performance of BONIK under different criteria. In Section \ref{sec:discussion}, we discuss how the design choices for BONIK have helped it to satisfy different requirements and explore its advantages, limitations and the possible future research scopes. Finally, we conclude in Section \ref{sec:conclusion}.

\section{Background}
\label{sec:back}
In this section we provide a brief background on blockchain technology (Section \ref{sec:back:subsec:bc}) and chatbots (Section \ref{sec:back:subsec:chatbot}). 
\subsection{Blockchain}
\label{sec:back:subsec:bc}
Bitcoin is regarded as the first widely-used decentralised digital currency that does not rely on a central entity, such as a central bank, for its creation and circulation \cite{nakamoto2019bitcoin}. Its main technological breakthrough is due to its underlying mechanism called \textit{blockchain}, an example of a distributed ledger shared among a group of Peer-to-Peer (P2P) nodes \cite{chowdhury2019comparative}. The ledger is an ordered data structure consisting of many blocks chained together by cryptographic mechanisms. Each block contains some transactions where each transaction enables a user to transact a certain amount of bitcoin to another user/users. Each block refers to its previous block using a cryptographic hash, which refers to its previous block and so on, hence forming a chain and colloquially known as \textit{blockchain}. 

Evolving from the Bitcoin blockchain, a new type of blockchain system has emerged, facilitating the deployment and autonomous execution of computer programs, known as \textit{smart-contracts}, on top of the respective ledger \cite{ferdous2019search}. Being part of the ledger makes smart-contracts and their executions immutable and irreversible, a sought-after property having a wide range of applications in different domains. Besides, a smart-contract supporting blockchain system has some other advantages, such as distributed data control, data persistence, data provenance, accountability, and transparency. Based on who can access a ledger in a blockchain system, there are generally two types of blockchain:
\begin{itemize}
    \item \textbf{Public blockchain:} In a public blockchain, also known as the \emph{permissionless blockchain}, anyone can join and participate in the network for blockchain governance and transaction creation at any time. Examples of public blockchain systems are Bitcoin \cite{bitcoin2018}, Ethereum \cite{ethereum2018}, Litecoin \cite{litecoin2011}, Monero \cite{monero2016} and so on. 
    \item \textbf{Private blockchain:} In a private blockchain, also known as \emph{permissioned} blockchain, only authorised and trusted entities are allowed to participate supporting different levels of permissions and privacy. Examples of private blockchain systems are Hyperledger Platforms \cite{hyperledger2018}, Quoram \cite{quorum2018}, and others. 
\end{itemize}

\subsection{Chatbot}
\label{sec:back:subsec:chatbot}
Chatbot (or a \textit{bot} in short) is an application program that can make auditory or textual conversations in real time with users \cite{abdul2015survey}. This is a smart implementation of AI providing a user-friendly conversational experience for users via multiple channels. It is the upcoming leading technology for vast potential for sales, customer service and marketing. In the next section, we explore several aspects of a chatbot.

\vspace{1mm}
\noindent \textbf{Use-cases.} Chatbots are increasingly being used as personal assistants for users, enabling people to converse with a chatbot, ask questions and get things done such as call someone, pay bills, set up a meeting and carry on many other activities that a personal assistant is supposed to do. On March 24, 2017, a 4 years old child \textit{Roman} even saved his mother's life using Siri, a chatbot from Apple \cite{siri2020, roman}. Other popular such chatbots are Google  Assistant \cite{googleAssistant2020} and Amazon's Alexa \cite{alexa2020}. Chatbots are also being used at call centres enabling customers to query regarding their products and receive instant replies $24/7$.

\vspace{1mm}
\noindent \textbf{Classification:} Bots can be classified mainly in two types \cite{classification}:
\begin{itemize}
    \item \textbf{Text-based:} A user interacts with a text-based chatbot with texts only. Users will query with texts and get answers with texts also. Such chatbots can be of two types. One is a bot providing fixed options and users need to select an option to interact with. The other is a dynamic chatbot where the bot, on taking random queries from a user, provides a dynamic answer to the user. 
    \item \textbf{Voice-activated:} This is the most sophisticated class of chatbots in which users interact with the bot using voice. 
\end{itemize}
\vspace{1mm}
\noindent \textbf{Mechanisms:} Here, we provide a simple working mechanism of a chatbot. A chatbot consists of a number of components. The front-facing component for a text chatbot is the User Interface (UI) using which a user interacts and submits queries or selects options. A voice-activated chatbot utilises the microphone of the corresponding devices to receive instructions/inputs from the user. An option-based text chatbot is the easiest to develop as it just needs to be equipped to handle a limited number of pre-selected options. Dynamic textual and voice chatbots, on the other hand, need to utilise a number of additional components and advanced algorithms, such as voice translation and Speech To Text Reporter (STTR), to function properly. These chatbots also need to apply other Natural Language Processing mechanisms, such as Part-Of-Speech Tagging \cite{brill1995transformation} and Sentiment Analysis \cite{bakshi2016opinion} to understand the query and to produce a a suitable output. 

\vspace{2mm}
\noindent\textbf{Financial chatbots:} A financial chatbot is a specific type of chatbot which is used in financial domains with a wide-range of use-cases, such as allowing users to execute financial transactions, providing financial advises, preventing financial frauds and so on \cite{okuda2018ai, finChatbot}. In the scope of this paper, we restrict out attention only to executing financial transactions. 

\vspace{2mm}
\noindent\textbf{Security and Privacy issues:} Because of their wide usages in different applications domains, chatbots often need to handle sensitive data. Therefore, the security and privacy issues are of great importance for chatbots. Here, we highlight a few of such issues, mostly applicable to financial chatbots, such as \textit{secure authentication, data confidentiality and integrity, system availability, accountability} and \textit{transparency} \cite{bozic2018security, lai2018banking, yan2018identifying}. Only authenticated users should be allowed to interact with a chatbot so that they can submit queries/transact for their respective bank account. Data confidentiality and integrity will guarantee that the submitted transaction is accessible by an authorised entity and is secure against any corruption. System availability will ensure uninterrupted access while accountability and transparency of the system will help to increase the trustworthiness of the system. The principal data privacy issues mostly arise from the lack of control and consent over any submitted transaction.

\section{Threat Modelling \& Requirement Analysis}
\label{sec:proposal}
In this section, we present a threat model (Section \ref{sec:proposal:subsec:threat}) and analyse a number of functional, security and privacy requirements (Section \ref{sec:proposal:subsec:req}) for a blockchain empowered financial chatbot.
\subsection{Threat Modelling}
\label{sec:proposal:subsec:threat}
Threat  modelling is an integrated process of designing a secure system which is used to identify, communicate, and understand threats and mitigation mechanisms within the context of protecting (IT) assets, financial transactions and chatbot in the scope of this paper. To model threats, we have chosen a well established threat model called STRIDE \cite{shostack2014threat}, developed by Microsoft, which encapsulate different security threats as presented below.

\begin{itemize}
    \item \textbf{T1-Spoofing Identity:} The act of spoofing refers to an adversary using the identity of an authorised user (e.g. a sender or a receiver of a financial transaction) to illegally access or participate in financial transactions.
    
    \item \textbf{T2-Tampering with Data:} An attacker can try to change a transacting amount in a financial transaction.
    
   \item \textbf{T3-Repudiation:} An attacker can repudiate certain invalid and illegal actions involving a financial transaction.

    \item \textbf{T4-Information Disclosure:} Sensitive data stored in the system is leaked to an attacker unintentionally.
    
    \item \textbf{T5-Denial of Service (DoS): } The system that is used to access the service can be the target of a DoS attack.
    
     \item \textbf{T6-Elevation of Privilege: } An attacker might use other attack vectors such as malicious software with potential exploitable vulnerabilities in order to execute transactions without the knowledge of a valid user.
\end{itemize}

In addition to these, we have considered an additional threat which is crucial for any financial system. 

\begin{itemize}
    \item \textbf{T7-Replaying Transactions:} An attacker might capture an old transaction and submit it afterwards, thus launching a replay attack.
\end{itemize}

The privacy threats mostly emerge from the lack of any privacy control by any user. Based on this assumption, the identified threats are as follows.

\begin{itemize}
 \item \textbf{T8-Lack of Consent:} A transaction is being carried out without the consent of a user.
 \item \textbf{T9-Lack of control and Transparency:} Users have little control on the way transaction is being carried out.

\end{itemize}
\subsection{Requirement analysis}
\label{sec:proposal:subsec:req}
In this section, we present a set of functional, security and privacy requirements. The functional requirements capture the core functionalities of the system while security and privacy requirements ensure that they mitigate the identified threats.

\vspace{2mm}
\noindent\textbf{Functional Requirements (FR):} The requirements are presented below.
\begin{enumerate}
    \item[F1.] Users should be able to execute financial transactions, e.g. balance query and transfer money, through the chatbot easily by interacting with it.
    \item[F2.] The chatbot should be integrated with a private blockchain infrastructure simulating banking functionalities so that financial transactions can be carried out without any error.
    \item[F3.] The system should ensure the transparency of the transactional data so that an authorised user can inspect different transactions when required, e.g. during a dispute.
\end{enumerate}

\noindent\textbf{Security Requirements (SR):} Next, we present a set of security requirements to address the identified security threats.
\begin{itemize}
    \item[S1.] The system must ensure that only securely authenticated users can avail this service. 
    \item[S2.] The system must ensure that one user's chatting information is not shared with another user. \textit{S1} and \textit{S2} can combinedly mitigate \textit{T1} threat.
    \item[S3.] Any conversational and transactional data must be transferred via networks in a secure manner so as to ensure the confidentiality, integrity and authenticity of a user's transaction data. This can mitigate \textit{T2}, \textit{T3} and \textit{T4} threats.
    \item[S4.] The system must guard against any DoS attack so as to mitigate the \textit{T5} threat.
    \item[S5.] The system must take protective measures against any replay attack in order to mitigate the \textit{T7} threat.

\end{itemize}

\noindent\textbf{Privacy Requirements (PR):} Privacy requirements are important to mostly mitigate the privacy threats. We present these requirements below.
\begin{itemize}
    \item[P1.] The system must ensure that each transaction activity must be carried out only with the user's consent. This mitigates \textit{T6} and \textit{T8} threats. 
    \item[P2.] The system should ensure that a user has full control over any of their transactions. This mitigates \textit{T9} threat.
\end{itemize}

\section{Architecture \& Implementation}
\label{sec:archi}

In order to effectively tackle the identified security and privacy issues involving a financial chatbot (as presented in Section \ref{sec:back:subsec:chatbot}), we propose to develop a chatbot rooted on a blockchain system. A blockchain system is decentralised in nature offering a secure transaction and time-stamping recording mechanism with a strong support for integrity and immutability. Moreover, a smart-contract empowered blockchain system offers the opportunity to deploy complex and immutable logic within a blockchain which can be invoked autonomously using transactions. Towards this aim, we present \textit{BONIK}, a blockchain empowered financial chatbot, in this paper.

\begin{figure}[h]
\includegraphics[scale=0.53]{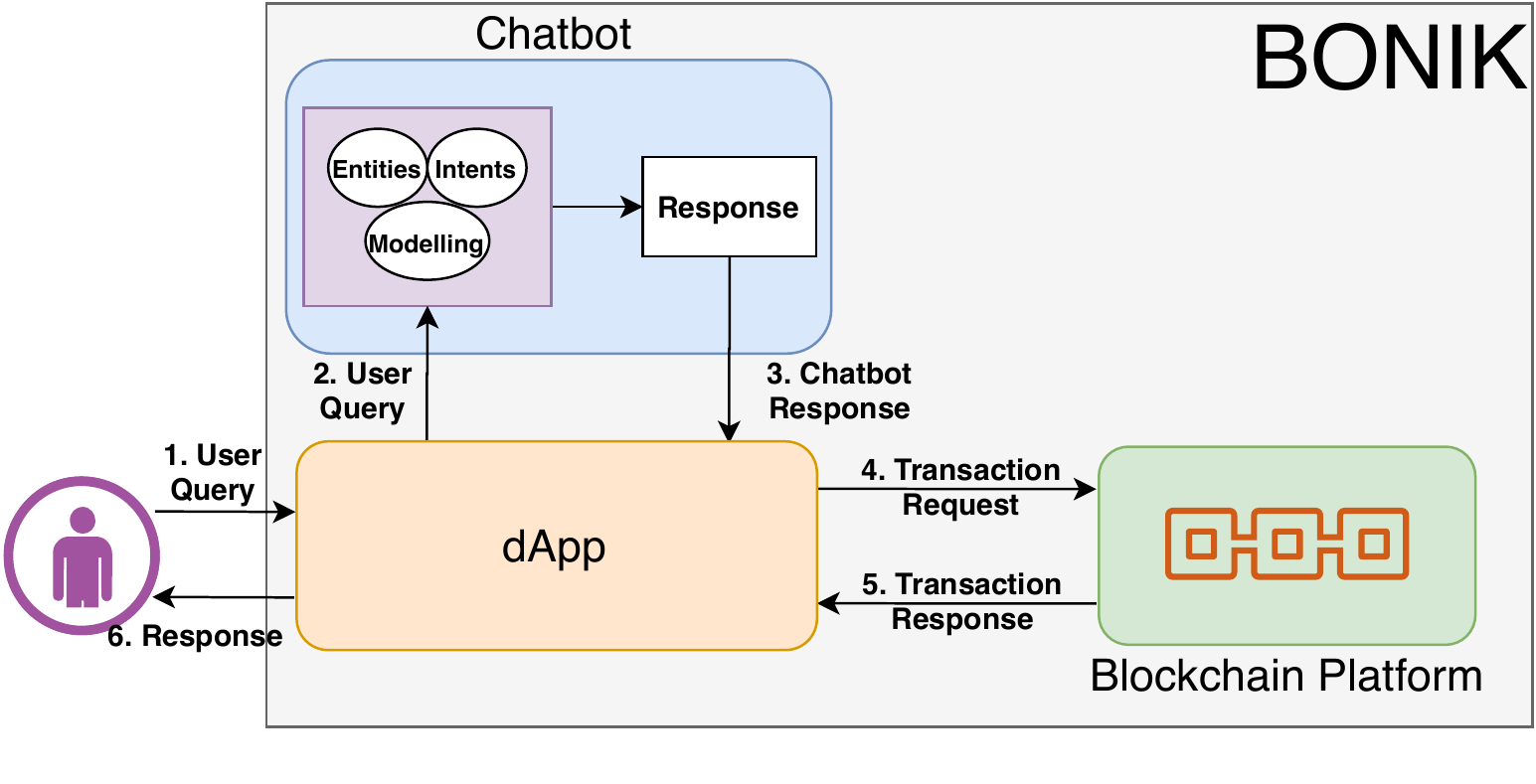}
\centering
\caption{Top-level architecture and flow in BONIK.}
\label{Fig:AppDialogflow}
\vspace{-2mm}
\end{figure}

A user can interact with BONIK to securely submit transactions and carry out financial activities such as querying for current balance. The blockchain integration enables BONIK to validate each request against pre-defined access control rules codified in smart-contracts and if only validated, user requests are honoured. We illustrate the top-level architecture of BONIK in Figure \ref{Fig:AppDialogflow}. This architecture consists of three main components, namely Chatbot, dApp (Decentralised Application) and the Blockchain platform. Next, we discuss the functionalities of each of these components along with their implementation details and inter-component interactions.

\subsection{Blockchain platform}
The chatbot in BONIK is integrated with a blockchain platform to facilitate a number of security features and some crucial functionalities. For example, the smart-contract in the platform simulates the functionalities of a financial institution such as a bank. Every user of the system is assumed to hold an account with this bank and every financial activity in the system, such as balance query and transfer money, is carried out by this bank. In the system, there are two smart-contracts providing business logic to handle user requests. The first one is provided by the system which handles user registration and login while the other is provided by the bank encoding the business logic for financial transactions. This compartmentalisation of smart-contracts provides modularity when other banks are added in the network, as they will just need to deploy and maintain their own smart-contracts without making much modifications in the system smart-contract.

For deploying the blockchain platform, we have studied different public and private blockchain systems. We have found that public blockchain systems are more secure, however, they are extremely slow, open to all and incur significant amount of cost to process and store data in a smart-contract supported public blockchain (e.g. Ethereum). Because of these reasons, we have chosen to work with a private blockchain system. Currently, Hyperledger Fabric is the most stable and popular private blockchain platform supporting smart-contract facility \cite{HyperledgerFabric}. It also provides a unique concept of \textit{channel} by which different blockchains can be maintained within the same network, thus creating a layer of privacy between different organisations, a must-have feature in any financial setting so that different activities remain private between different organisations. That this why we have selected to use Hyperledger Fabric as our preferred blockchain system during deployment.

\begin{figure}[h]
\centering
\includegraphics[scale=0.52]{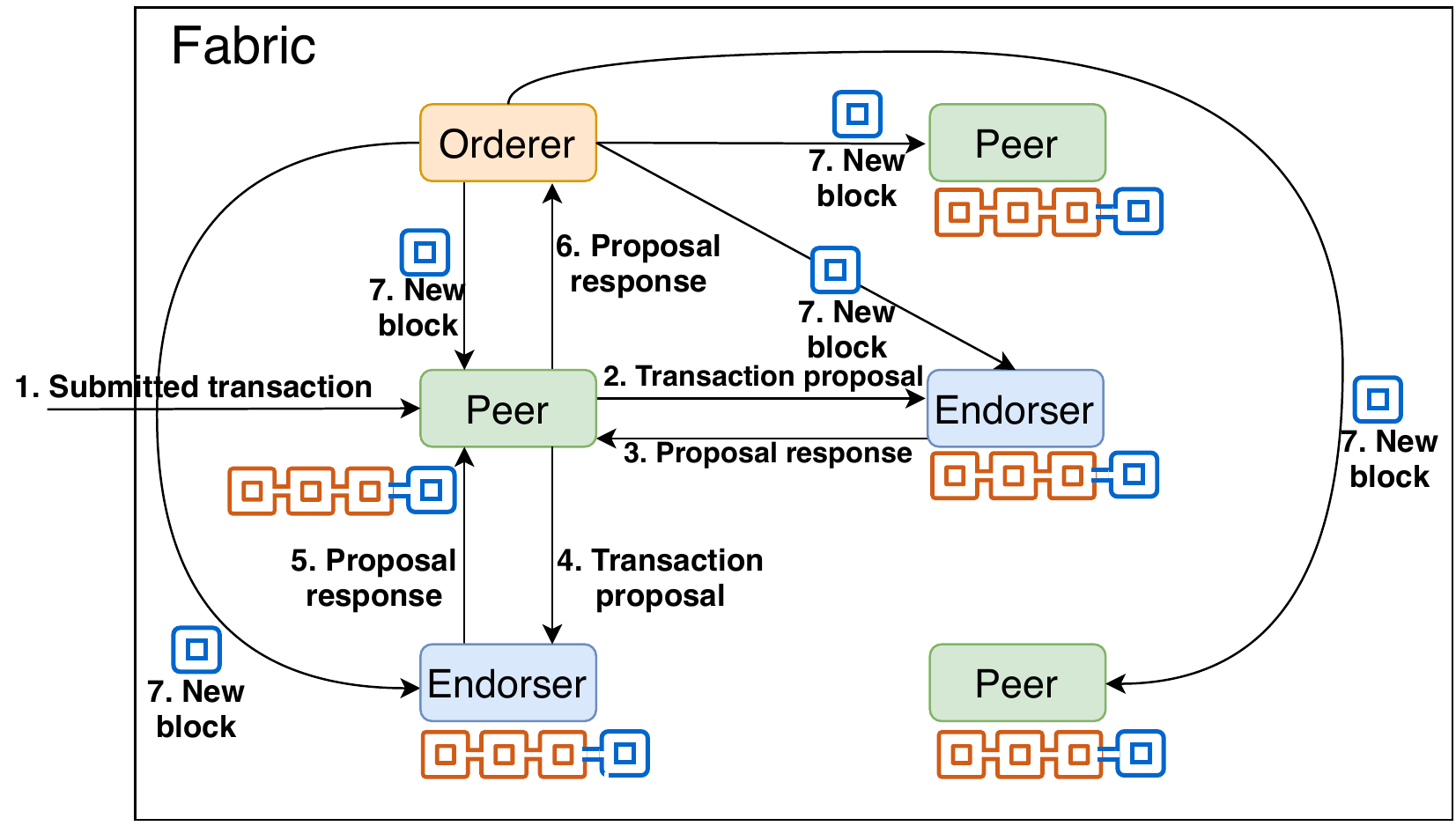}
\caption{Flow of activities in Fabric}
\label{Fig:Fabric}
\vspace{-1mm}
\end{figure}

Fabric utilises a number of network entities such as peers, endorsers and orderers (Figure \ref{Fig:Fabric}). A smart-contract is called a \textit{chaincode} in Fabric terminology which can be invoked using transactions. A user utilises a peer for submitting a transaction which is forwarded to the endorser(s) (steps 1, 2 \& 4 in Figure \ref{Fig:Fabric}). Each endorser is responsible for validating a transaction by checking if an entity is allowed to perform a certain action in a ledger encoded within the transaction (steps 3 \& 5 in Figure \ref{Fig:Fabric}). The validated transaction is then forwarded to the orderer(s). The Orderer creates a block using the transaction and returns the block to the endorsers and peers which is then added to the blockchain and thus, updating the state of the ledger (steps 6 \& 7 in Figure \ref{Fig:Fabric}). Consequently, a response is returned to the user. All the  entities (peers, orderers and the endorsers) are registered and authenticated via a Fabric specific special entity called \textit{Membership Service Provider} (MSP). This ensures that only authorised entities are allowed in the blockchain network.

The chaincode for BONIK has been written in Go where the blockchain platform consists of two organisations (representing the blockchain authority and the bank), where each organisation consists of two endorsers and peers. The blockchain platform is deployed using Docker containers where each container assumes the role of one of these entities. In addition, there is an additional container which plays the role of the MSP including the CA (Certificate Authority). These entities are connected via a channel into which the chaincode is deployed. The consensus is based on Kafka which utilises two additional orderer nodes for block creation and dissemination as described above. During the starting phase, each user is initialised with 10,000 unit of currency for executing financial transactions.

\subsection{dApp}
A chatbot is essentially a web application, however, has no mechanism to interact with a blockchain platform. To breeze this gap, we need a dApp which creates an interface between these two. A dApp (Decentralised Application) is configured as a web server exposing APIs to web applications and is also connected to a peer of the blockchain platform. Web applications use these APIs to submit queries which are translated into blockchain transactions by the dApp. Then, the dApp uses the blockchain API to submit these transactions, via a peer, for invoking a smart-contract on the blockchain and then the usual flows, as illustrated in Figure \ref{Fig:Fabric}, takes place.

The dApp in BONIK has been developed using Node.js with Express \cite{nodejs, express}. Node.js is a server-side JavaScript platform that is widely used for creating dApps in the blockchain domain. Express is a web application framework for Node.js which is used for developing web applications. Fabric provides the required APIs to interact with any Node.js application. We have developed the dApp in such a way that it is integrated with the chatbot and Fabric APIs and can facilitate the flow described above.

\subsection{Chatbot}
Chatbot is the component with which a user interacts, via their browser, to submit different financial queries. BONIK utilises a text-based dynamic chatbot enabling the user to generate random queries and receive the corresponding responses. We have developed chatbot using \textit{Dialogflow} agent that uses Google's  strong Natural Language Understanding (NLU) \cite{dialogflow} modules in the backend. It functions by taking inputs from a user as a query, processing the data after training itself using machine learning techniques and giving a response in return.

There are three basic elements in Dialogflow: \textit{intents}, \textit{entities} and \textit{training}. An intent refers to the mapping between the user's query and the agent's response. Each intent looks like a cluster where seemingly different queries map to a single preset output matched with a high probability. Indeed, a user may ask the system to initiate a transaction in many ways, however, the agent should detect that this is a user's attempt to create a transaction and responds with a single output.  

Entities in Dialogflow extract the parameter values from a user's input with natural language. For example, with respect to BONIK, a query for a transaction (e.g. \say{\textit{send account no 1123158964 1000 unit}}) will have several  corresponding entities: account number (1123158964) and amount (1000 unit). Training enables the Dialogflow agent to understand what user implies and approaches them in a structured way. Machine Learning techniques are used in the backend for this purpose which enable the Dialogflow agent to cluster similar intents and handle entities. For this, Dialogflow uses their own language models. In addition, a developer can feed in their own training data suitable for a particular application. Based on these two, Dialogflow trains itself to handle user queries and generates responses. This model improves dynamically as users converse more and more with the agent which increases its performance and reliability. For our system, a Dialogflow agent named 
`Transactional Chatbot' has been created. dApp interacts with this Dialogflow agent by calling the corresponding API with necessary information. The ML model of the agent has been trained with two datasets, namely user dataset and bot dataset, both have been developed by us for BONIK. The user dataset consists of the set of queries that a user can generate. The bot dataset, on the other hand, consists of sentences which are generated in response to any query from the user dataset.

Next, a high-level flow involving different components of BONIK is discussed. Once the user is securely logged in (the process is described in the subsequent section), the user can submit different queries to the dApp via the Chatbot UI. These queries are passed to the Dialogflow service and are handled accordingly. When the response is returned from Dialogflow, the dApp processes and parses the response. If additional query is required, the response is returned to the user via the UI. If the response is sufficient to create a Fabric transaction, the dApp converts that response into a transaction which is then submitted to the blockchain platform via the connected peer and then the usual flows take place. If the transaction is for balance query, upon receiving the response from the chaincode, the dApp displays the result on the UI. If the transaction is for transferring funds and the transfer is successful, an appropriate message is shown to the user. Alternatively, an appropriate error message is shown to the user via the UI if there is any error executing the transaction. The dApp flow in BONIK is illustrated in Figure \ref{Fig:AppDialogflow}.

\section{Protocol Flow}
\label{sec:protocol}
In this section, we present the protocol flow between different components in BONIK. Before we illustrate the protocol flow, we introduce mathematical notations in Table \ref{table:cryNot} and data model in Table \ref{table:dModel}.

\begin{table}[t]
\caption{Cryptographic Notations}
\label{table:cryNot}
\centering
\begin{tabular}{r|l}
\hline
\textbf{Notations}            & \textbf{Description} \\ \hline \hline

$K_{U_f}$      & Public key of the sender.  \\ \hline
$K^{-1}_{U_f}$      & Private key of the sender.  \\ \hline
$K_d$      & Public key of the Dapp.  \\ \hline
$N_i$    & A fresh nonce.   \\ \hline
${\{\}}_K$     & Encryption operation using a public key $K$.  \\ \hline
${\{\}}_{K^{-1}}$     & Signature using a private key $K^{-1}$.  \\ \hline
$H(M)$     & SHA-256 hashing operation of message $M$.  \\ \hline
${[ ]}_{\mathit{https}}$     & Communication over HTTPS channel $K$.  \\ \hline
\end{tabular}
\end{table}

\begin{table}[h]
\caption{Data Model}
\label{table:dModel}
\centering
\begin{tabular}{l}
\hline
$\mathit{req} \define \langle type, data \rangle$\\ \hline
$\mathit{resp'} \define \langle resp, K_{U_f}, K^{-1}_{U_f} \rangle$\\ \hline
$\mathit{TYPE} \define \langle \mathit{registration, login, balQuery, transfer} \rangle$\\ \hline
$\mathit{DATA} \define \langle \mathit{regisData, loginData, balData, transferData} \rangle$\\ \hline
$\mathit{regisData} \define \langle \mathit{userName, h} \rangle$\\ \hline
$\mathit{loginData} \define \langle \mathit{userName, h} \rangle$ \\ \hline
$\mathit{balData} \define \langle \mathit{userName, accountNum} \rangle$ \\ \hline
$\mathit{transferData} \define \langle \mathit{userName, fromAcc, toAcc, amount} \rangle$ \\ \hline
$\mathit{string} \define \langle \mathit{string}_1, \mathit{string}_2, ..., \mathit{string}_n \rangle$\\ \hline
\end{tabular}
\end{table}
\noindent \textbf{Data Model:} We start with the request (denoted with $\mathit{req}$ in Table \ref{table:dModel}), which is submitted to the blockchain platform. $\mathit{req}$ consists of $\mathit{type}$ and $\mathit{data}$. Here, $\mathit{TYPE}$ denotes the set of different data types within a request and $\mathit{type} \in \mathit{TYPE}$ whereas, $\mathit{DATA}$ represent the set of corresponding data and $\mathit{data} \in \mathit{DATA}$. Both $\mathit{TYPE}$ and $\mathit{DATA}$ are defined as presented in Table \ref{table:dModel}. 

Next, $\mathit{registration}$ in type signifies that the corresponding request will be a registration request consisting of the data set denoted with $\mathit{regisData}$ and so on. In $\mathit{regisData}$, $h = H(Password)$ denotes the hash of the provided password and $\mathit{userName}$ denotes the username (identifier) of the user. This implies that a registration request must contain a username and the hash of the password. $\mathit{loginData}$ also has the similar semantic in the sense that a login request must consist of the username and the hash of the provided password. 

$\mathit{balData}$, on the other hand, is used for balance query and consists of the $\mathit{userName}$ and $\mathit{accountNum}$, implying that it must provide the username of the user and the account number to retrieve the balance of the user. Finally, $\mathit{transferData}$ is used for balance transfer requiring the username of the user as well as the sender's account number ($\mathit{fromAcc}$), the receiver's account number ($\mathit{toAcc}$) and amount to transfer ($\mathit{amount}$).

Next, we model the functionality of Dialogflow in which a user query is submitted and a set of entities is returned. A user query, in essence, represents the interactions between the user and the chatbot for a meaningful request. For example, a balance transfer query will consist of all required interactions between the user and the chatbot. We use the notations $\mathit{STRING}$, $\mathit{ENTITY}$ to denote the sets of strings (representing a Dialogflow interaction) and entities (as generated by Diaglogflow algorithm). Next, we define a function to model the core functionality of Dialogflow: transforming a string of query into a set of entities.

\begin{definition}\label{dflow-def}
Let $\mathit{dFlowModel} : \mathit{string} \rightarrow \mathcal{E}$ be the function that transforms a string into a set of entities.
\end{definition}

Here, $\mathit{string} \subseteq \mathit{STRING}$ and $\mathcal{E} \subseteq ENTITY$. In other words, $\mathit{string}$ represents the set of all elements from the user and chatbot datasets required to build a meaningful balance query and balance transfer query.  $\mathit{string}$ is modelled as presented in Table \ref{table:dModel}, where $\mathit{string}_1, \mathit{string}_2, ..., \mathit{string}_n$ represent different elements from the user and chatbot datasets as submitted by the user and the chatbot while interacting for a particular request.

The dApp in BONIK is responsible for handling the returned set of entities ($\mathcal{E}$) which is parsed into corresponding requests, either a balance request or transfer request. We define the following function to model this parsing capability.

\begin{definition}\label{parsing-def}
Let $\mathit{parsing} : \mathcal{E} \rightarrow \mathit{req} $ be the function that transforms a set of entities into a corresponding request.
\end{definition}
\vspace{2mm}
\noindent \textbf{Algorithms:} We present the algorithms of the system chaincode and bank chaincode in Algorithm \ref{algo:scc} and Algorithm \ref{algo:bcc} respectively.

\begin{algorithm}
\SetAlgoLined

\caption{SCC:\,\,\,/\,\,/\,\,\,\,\,\texttt{$\triangleright$ System Chaincode}}

\label{algo:scc}
\textbf{Input:} $\mathit{req} \rightarrow$ the request from the user \\
\textbf{Output:} $\mathit{resp} \rightarrow$ the chaincode generated response\\
\SetKwBlock{Begin}{}{}
\Begin(\textbf{Start})
{

  \SetKwFunction{invoke}{\textbf{invoke}}
  \SetKwProg{Fn}{function}{}{}
  \Fn{\invoke{$\mathit{req}$}}{
    $\mathit{data} := \mathit{req}.data$\;
    $\mathit{type} := \mathit{req}.type$\;
    \uIf{req.type == login}{
        $\mathit{resp} = $ loginFunc($\mathit{data}$)\; 
    }
    \uElseIf{req.type == registration}{
        $\mathit{resp} = $ regFunc($\mathit{data}$)\; 
    }
    \Else{
        $\mathit{resp} = $BankCC.invoke($\mathit{req}$)\; 
    }
    send $\mathit{resp}$ back to user\;
  }

  \SetKwFunction{reg}{\textbf{regFunc}}
  \SetKwProg{Fn}{function}{}{}
  \Fn{\reg{$\mathit{data}$}}{
    $\mathit{uName} := data.userName$\;
    $\mathit{h} := data.h$\;
    $\mathit{putState(uName,h)};$ $\triangleright${ store into blockchain}\\
    \KwRet $\mathit{TRUE}$\;
  }

  \SetKwFunction{login}{\textbf{loginFunc}}
  \SetKwProg{Fn}{function}{}{}
  \Fn{\login{$\mathit{data}$}}{
        $\mathit{uName} := data.userName$\;
        $\mathit{hPasswd} = \mathit{getState(uName)};$ $\triangleright${ retrieve from blockchain}\\
        \uIf{data.h == $\mathit{hPasswd}$}{
            \KwRet $\mathit{TRUE}$\;
        }
        \Else{
            \KwRet $\mathit{FALSE}$\; 
        }
  }
}
\end{algorithm}

\begin{algorithm}
\SetAlgoLined

\caption{BCC:\,\,\,/\,\,/\,\,\,\,\,\texttt{$\triangleright$ Bank Chaincode}}
\label{algo:bcc}
\textbf{Input:} $\mathit{req} \rightarrow$ the request from the user \\
\textbf{Output:} $\mathit{resp} \rightarrow$ the chaincode generated response\\
\SetKwBlock{Begin}{}{}
\Begin(\textbf{Start})
{

  \SetKwFunction{invoke}{\textbf{invoke}}
  \SetKwProg{Fn}{function}{}{}
  \Fn{\invoke{$\mathit{req}$}}{
    $\mathit{data} := \mathit{req}.data$\;
    $\mathit{type} := \mathit{req}.type$\;
    \uIf{req.type == balQuery}{
        $\mathit{resp} = $ balQFunc($\mathit{data}$)\; 
    }
    \Else{
        $\mathit{resp} = $ transFunc($\mathit{data}$)\; 
    }
    send $\mathit{resp}$ back to SCC\;
  }

  \SetKwFunction{balq}{\textbf{balQFunc}}
  \SetKwProg{Fn}{function}{}{}
  \Fn{\balq{$\mathit{data}$}}{
    $\mathit{uName} := data.userName$\;
    $\mathit{acct} := data.accountNum$\;
    $\mathit{balance} = \mathit{getState(acct)}$\; 
    \KwRet $\mathit{balance}$\;
  }

  \SetKwFunction{trans}{\textbf{transFunc}}
  \SetKwProg{Fn}{function}{}{}
  \Fn{\trans{$\mathit{data}$}}{
    $\mathit{uName} := data.userName$\;
    $\mathit{fromAcct} := data.fromAcc$\;
    $\mathit{toAcct} := data.toAcc$\;
    $\mathit{amount} := data.amount$\;
    $\mathit{fromBalance} = \mathit{getState(fromAcct)}$\;
    $\mathit{toBalance} = \mathit{getState(toAcct)}$\; 
    \uIf{$\mathit{fromBalance} > \mathit{amount}$}{
        $\mathit{fromBalance} \mathrel{{-}{=}} \mathit{amount}$\;
        $\mathit{toBalance} \mathrel{{+}{=}} \mathit{amount}$\;
        $\mathit{putState(fromAcct,fromBalance)}$\; 
        $\mathit{putState(toAcct,toBalance)}$;\; 
        \KwRet $\mbox{``TRANSACTION SUCCESSFUL"}$\;
    }
    \Else{
        \KwRet $\mbox{``TRANSACTION ABORTED"}$\;
    }
  }
 } 
\end{algorithm}

Whenever the system chaincode (represented as $\mathit{SCC}$ in Algorithm \ref{algo:scc}) receives a request (denoted with $\mathit{req}$ in the algorithm), its invoke function is initiated. This function retrieves $\mathit{data}$ and $\mathit{type}$ from the request (line 5 and 6) and then invokes any of the other two functions, \textit{regFunc} and \textit{loginFunc}, depending on the request type (line 7 to 10). For example, the \textit{loginFunc} encodes the logic for the login functionality whereas the \textit{regFunc} encodes the registration functionality. Once executed, a response is returned (denoted with $\mathit{resp}$) back to the dApp (line 14).

The bank chaincode (represented as \textit{BankCC} in Algorithm \ref{algo:bcc}) consists of three functions, namely \textit{invoke, balQFunc} and \textit{transFunc}. The \textit{balQFunc} encodes the algorithm for the balance query operation and the \textit{transFunc} encodes the logic for the balance transfer operation. When $\mathit{invoke}$ receives $\mathit{req}$, data and type values are retrieved from the request (line 5 and 6 in \ref{algo:bcc}). Depending on the type values, the corresponding function is called with data (line 7 to 10). After executing their code, each of these two functions return a result which is stored in the response (denoted with $\mathit{resp}$, line 12 in the algorithm) and is then returned back to the system chaincode which consequently returns the response back to dApp.

\begin{figure}[h]
\includegraphics[scale=0.65]{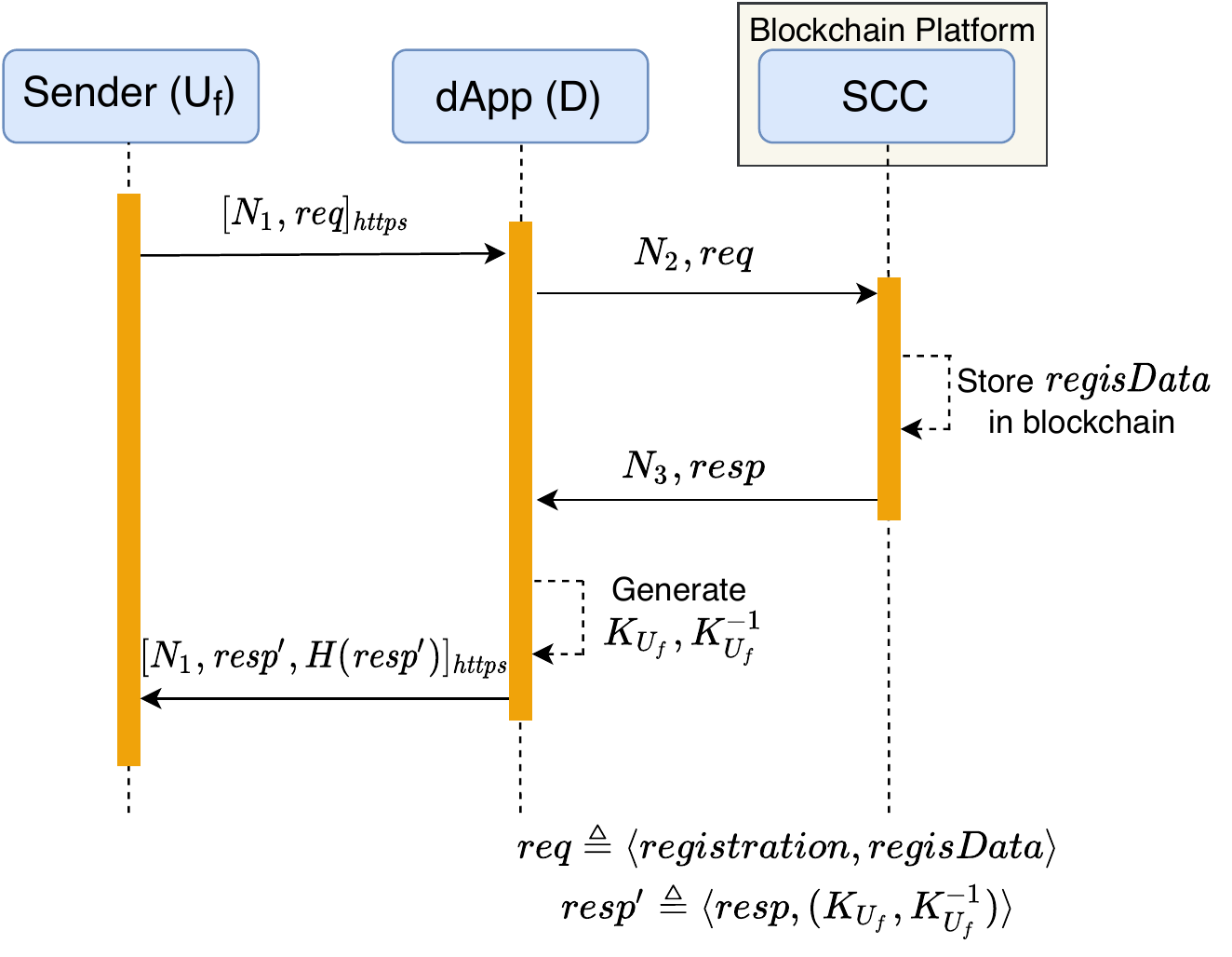}
\centering
\caption{Registration flow in BONIK.}
\label{Fig:RegisFlow}
\end{figure}

\noindent \textbf{Protocol flow:} Now, we present the protocol flow illustrating user interactions with different components in BONIK. To interact with BONIK, a user must register herself following the protocol presented in Table \ref{table:regProtocol} and illustrated in Figure \ref{Fig:RegisFlow}. Here, the user submits a username and password in the registration form. The password is hashed using SHA-256 hashing algorithm in the client side. This $\mathit{userName}$ and the hashed password make up $\mathit{regisData}$ where  $h$ denotes the hashed password. The $\mathit{req}$ in the registration process consists of $\mathit{registration}$ type and $\mathit{regisData}$. As per the protocol, in the first message (denoted with $M1$ in Table \ref{table:regProtocol}), a user ($U_f$) sends to the dApp a nonce ($N_1$), $\mathit{req}$ encrypted with the public key of the dApp ($K_d$), over an HTTPS channel. dApp decrypts the request using its private key and forwards this request to $\mathit{SCC}$ (M2 in Table \ref{table:regProtocol}). This is handled in the \textit{regFunction} where the username and the hashed password are extracted and are stored in the blockchain (line 16 to 18 in Algorithm \ref{algo:scc}). Then a $\mathit{TRUE}$ value is returned to the calling code (the $\mathit{resp}$ variable in line 10), signifying that the user registration response is successful. This response is returned to dApp. Next, dApp generates public and private keys for $U_f$ ($K_{U_f}$ and $K^{-1}_{U_f}$ respectively) using Fabric MSP functionality. This key pair and the response ($\mathit{resp}$) from $SCC$ are combined to create $\mathit{resp'}$ (see Figure \ref{Fig:RegisFlow}). Then, this response and its SHA-256 hash ($\mathit{resp'}, H(\mathit{resp'})$) are returned to $U_f$ over an HTTPS channel. Then, the user stores her public and private keys in her device for any future correspondence. 

\begin{table}[h]
\caption{Registration protocol}
\label{table:regProtocol}
\centering
\begin{tabular}{lll}
\hline
$M1$ & $U_f \rightarrow D:$  & $[N_1, {\{\mathit{req}\}}_{K_d}]_{\mathit{https}}$ \\
$M2$ & $D \rightarrow SCC:$  & $N_2, \mathit{req}$ \\
$M3$ & $SCC \rightarrow D:$  & $N_2, \mathit{resp}$ \\
$M4$ & $D \rightarrow U_f:$  & $[N_1, \mathit{resp'}, H(\mathit{resp'})]_{\mathit{https}}$\\
\hline
\end{tabular}
\end{table}

Every user must log in before accessing the service. The login protocol is similar to the registration protocol where the user submits the username and password via their browser. These data are encoded into an appropriate $\mathit{req}$ and submitted to dApp which invokes the $\mathit{loginFunc}$ in $\mathit{SCC}$ to handle this request (line 20 to 26 in Algorithm \ref{algo:scc}). A successful validation will sign in the user to the system. For security, every request and response between the user and the dApp are signed with the sender's private key and are transmitted over HTTPS.
\begin{figure*}[!h]
\includegraphics[scale=0.66]{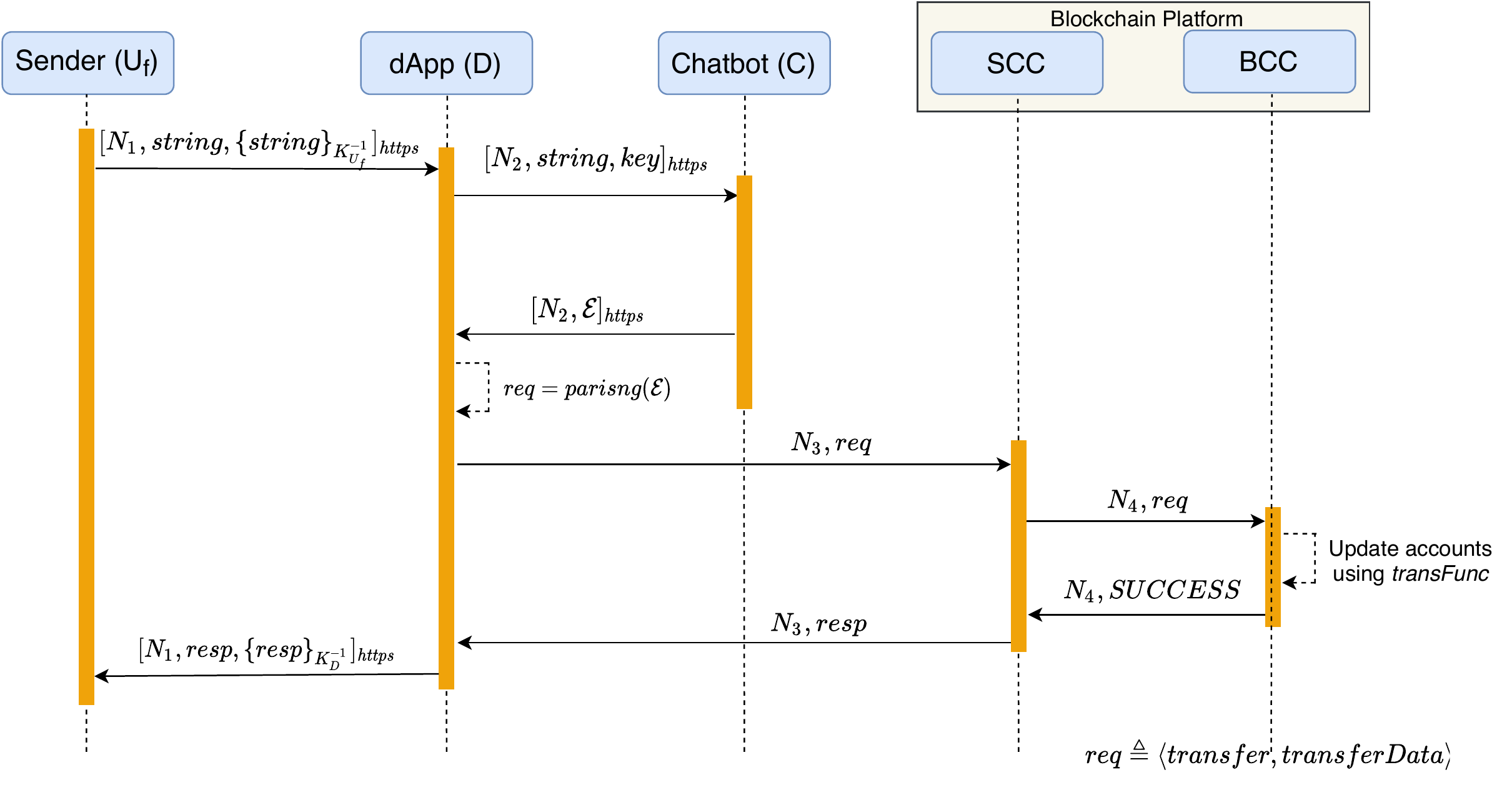}
\centering
\caption{Balance transfer flow in BONIK.}
\vspace{-5mm}
\label{Fig:TransferFlow}
\end{figure*}

Next, we present the protocol flow for the balance transfer from $U_f$, assuming $U_f$ is already logged in. The protocol is illustrated in Figure \ref{Fig:TransferFlow}. Once $U_f$ logs in, a chat interface is loaded in her web browser to interact with the chatbot. $U_f$ submits a query for balance transfer (denoted with $\mathit{string}$ Figure \ref{Fig:TransferFlow}), using this interface, to dApp along with a nonce. It is to be noted, as per the mathematical model, $\mathit{string}$ encodes an interaction between the user and the Dialogflow chatbot consisting of a number of texts required for a meaningful query. We have not shown this interaction in the protocol flow for brevity.

Like before, $\mathit{string}$ is signed with $K^{-1}_{U_f}$ and transmitted over an HTTPS channel. After a successful signature verification, this string along with a secret key (denoted with $\mathit{key}$ in Figure \ref{Fig:TransferFlow}) is forwarded to Dialogflow over an HTTPS channel. Every request submitted to Dialogflow API must be registered and authorised beforehand. The secret key is used to validate the authorisation. Then, Dialogflow utilises its $\mathit{dFlowModel}$ function to convert this string to a set of entities ($\mathcal{E}$) which is returned to dApp. dApp utilises its $\mathit{parsing}$ function to convert it to a balance transfer request (consisting of $\mathit{balQuery}$ and $\mathit{balData}$). dApp then invokes $SCC$ with this request which is internally forwarded to the invoke function of $BCC$. This balance transfer request consequently invokes the $\mathit{transFunc}$ (line 10 in Algorithm \ref{algo:bcc}) where the balances of the corresponding users' accounts are retrieved from the blockchain and after a validity check (if the user has sufficient balance), accounts are updated with the correct balance and stored in the blockchain (as outlined in line 19 to 29 in Algorithm \ref{algo:bcc}). A successful balance transfer operation will return a ``\textit{TRANSACTION SUCCESSFUL}'' response, otherwise a ``\textit{TRANSACTION ABORTED}'' response will be returned. This response will be returned back to dApp and from there ultimately to the user over HTTPS. The balance query protocol for $U_f$ will be similar and is excluded for brevity.
\section{Evaluation}
\label{sec:evaluation}
To evaluate the performance of BONIK, we have utilised Hyperledger Caliper \cite{caliper}, a state-of-the-art blockchain benchmarking tool for Hyperledger blockchain platforms, including Hyperledger Fabric. With BONIK integrated with Caliper, we can measure the performance of its blockchain implementation with a set of predefined network configurations such as the number of entities within the network, the number of simulated users and requests accessing BONIK simultaneously. 

The experiment has been carried out in a PC with a Ubuntu 18.04-64 OS and hardware configurations of Intel(R) Core i5-8265U @1.60GHz quad-core CPU, 8 GB DDR4 RAM, 256 GB SSD, 1 TB HDD and 2GB GeForce MX150 Graphics GPU. We have simulated between $10$ to $50$ users who have submitted different transactions for creating users (registrations), balance query and transfer at varied degrees of rate with three different network configurations consisting of 2 orderers 2 peers (denoted with 2O2P), 2 orderers 4 peers (2O4P) and 2 orderers 6 peers (2O6P).

Caliper supports a wide-range of different configurations. Before our main experiments, we have tested these configurations to identify the ideal setup which is the following. The amount of time to wait before creating a batch, the \textit{Batch Timeout} is set as 1s. The maximum message count for a single batch is set as 500 and the transaction rate is set 20 per second. With these configurations, each experiment has been carried out $5$ times and the result is then averaged and presented next.

\begin{figure*}[h]

  \begin{subfigure}{0.325\textwidth}
    \includegraphics[width=\linewidth]{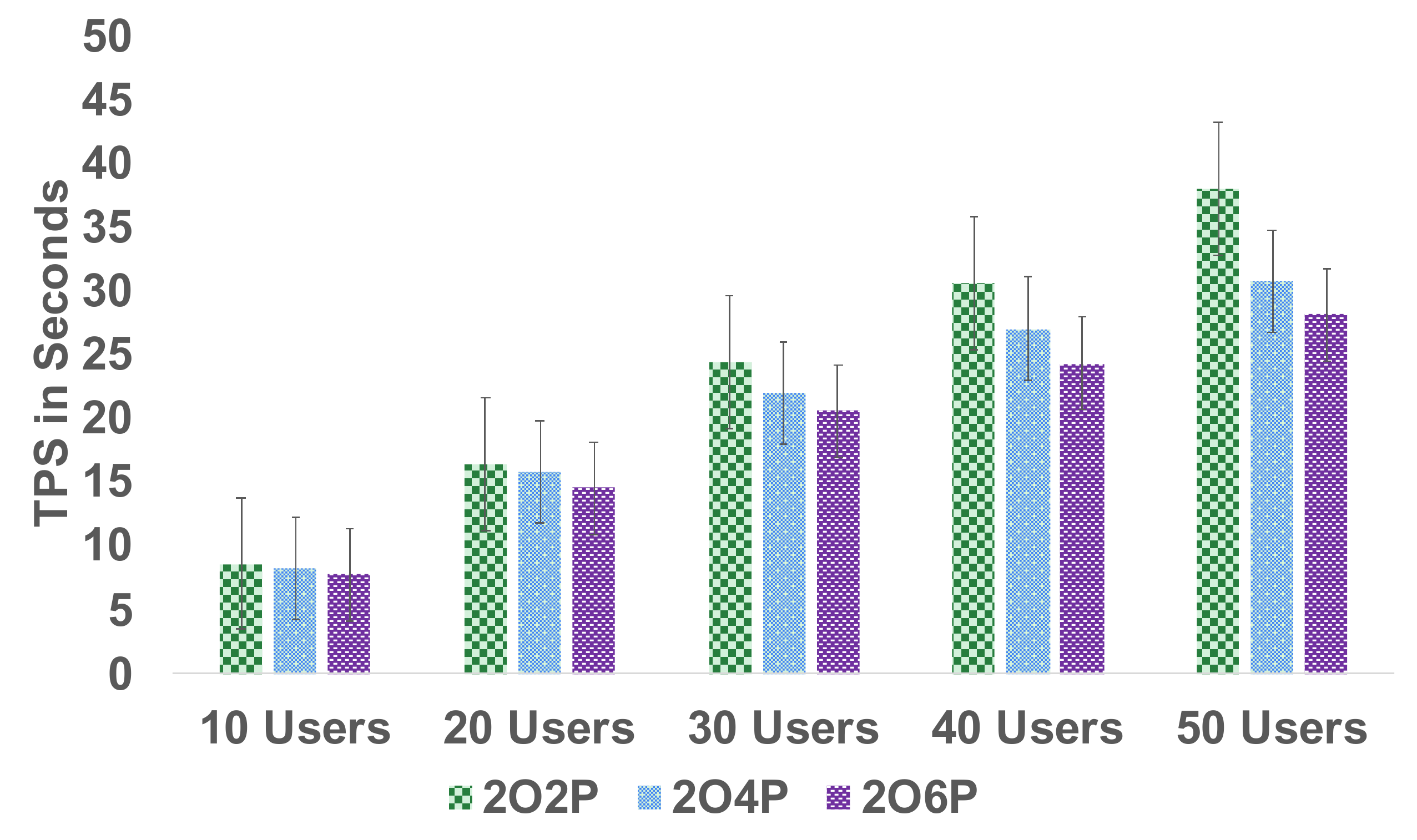}
    \caption{User creation} \label{fig:1a}
  \end{subfigure} 
  \hspace*{\fill}    
  \begin{subfigure}{0.325\textwidth}
    \includegraphics[width=\linewidth]{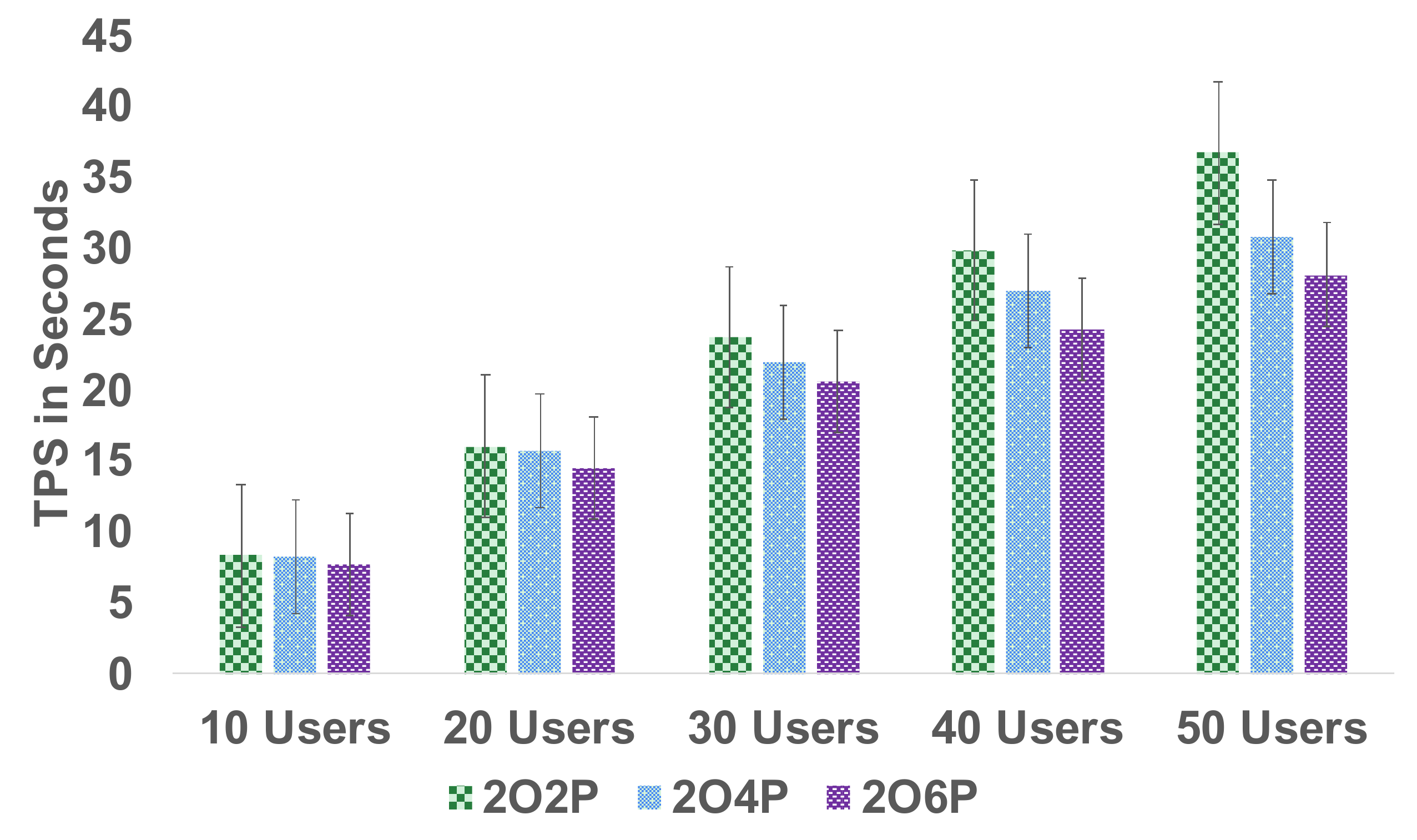}
    \caption{Balance transfer}  \label{fig:1b}
  \end{subfigure}
  \hspace*{\fill}   
  \begin{subfigure}{0.325\textwidth}
    \includegraphics[width=\linewidth]{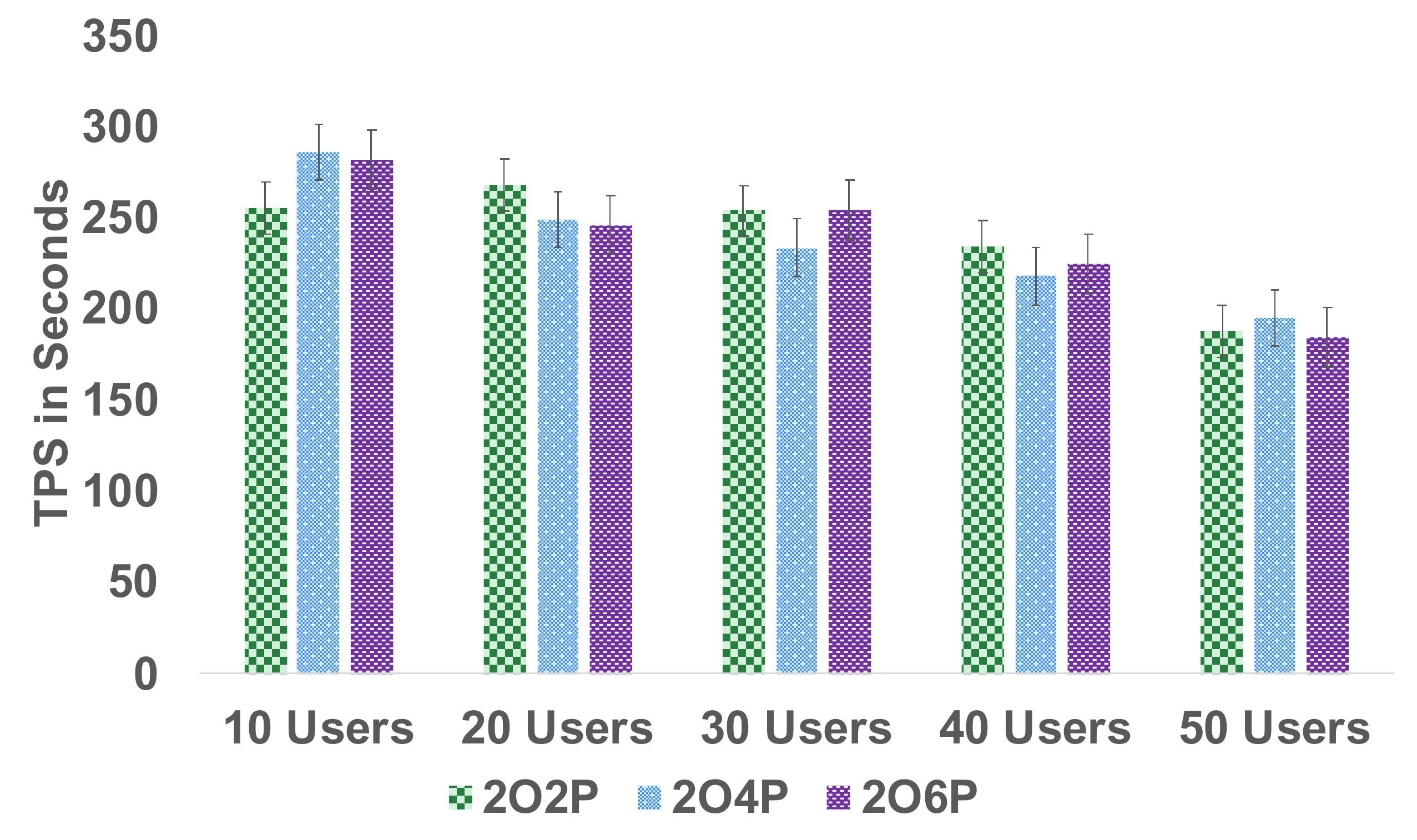}
    \caption{Balance query}  \label{fig:1c}
  \end{subfigure}
  \caption{Transaction Per Second (TPS) using BONIK} \label{fig:tps}
\vspace{-3mm}  
\end{figure*} 

\subsection{User Creation}
In Figure \ref{fig:1a} the average TPS vs different number of users against different configurations is plotted. Two trends are clear from this figure. The first trend is that TPS increases with the number of users in every configuration. For example, in the 2O2P configuration, the TPS for 10 users is $8.6$ which increases to $37.98$ for 50 users under the same configuration, a $4x$ increase. This trend is seen in other configuration sets as well. This seems counter-intuitive, however, the underlying reason for this increase is because of the batching mechanism in Fabric in which Fabric waits for a certain number of transaction for putting them in a single block. With more users, more transactions are batched together within a single block, thus resulting in higher TPS. The second trend is that TPS decreases within the same user set with increased number of entities. For example, in 50 users set, TPS decreses from $37.98$ to $28.14$ for 2O2p and 2O6P respectively. As the number of entities increases in Fabric network, it takes more time for endorsing and creating blocks, resulting in decreased TPS.

\subsection{Balance Transfer}
The performance for balance transfer experiment is presented in Figure \ref{fig:1b}. It exhibits similar treads, as in Figure \ref{fig:1a}, TPS increases as the number of users increases while, within the same user set, TPS decreases as the number of network entities in Fabric increases. Furthermore, in both experiments, TPS remains almost similar. For example, in 2O2P setting for 50 users, the TPS is $37.98$ and $36.72$ for user creation and balance transfer respectively.

\subsection{Balance Query}
The result for balance query is presented in Figure \ref{fig:1c}. With a maximum TPS of $286.16$ for 10 users in 2O2P setting, TPS for balance query is significantly higher than the previous two experiments. The main reason is that balance query is essentially a read operation from the chaincode which can be carried out locally from the Fabric, thus significantly reducing the latency and increasing the TPS. However, as the number of users increases, the TPS tends to to decrease for balance query as well: from $286.16$ for 10 users to $194.9$ for 50 users in the same 2O2P setting.

\section{Discussion}
\label{sec:discussion}
In this section, we examine how BONIK has satisfied its different requirements (Section \ref{sec:discussion:subsec:req}), discuss its advantages and limitations (Section \ref{sec:discussion:subsec:ad}) and highlight possible future works (Section \ref{sec:discussion:subsec:future}). 

\subsection{Analysing Requirements}
\label{sec:discussion:subsec:req}
Here, we explore if BONIK satisfies the formulated requirements of Section \ref{sec:proposal:subsec:req}. 

\vspace{1mm}
\noindent\textbf{Functional Requirements:} BONIK enables a user to submit financial transactions for balance query and transfer to another account, thereby satisfying \textit{F1}. The blockchain component in BONIK is based on Hyperledger Fabric, a private blockchain platform. The bank chaincode within the platform simulates the banking functionalities in an immutable and error-free manner, and hence, BONIK satisfies \textit{F2}. BONIK, underpinned by Fabric, inherits a core property of any blockchain platform, transparency, which enables every authorised entity to validate and verify any transaction. Thus, BONIK satisfies \textit{F3}.

\vspace{1mm}
\noindent\textbf{Security Requirements:} The BONIK protocol requires every user to be registered and authenticated before submitting any financial transactions. This ensures \textit{S1}. A secure session is maintained for each logged in user to create a layer of separation between different users. In this way, no user can access another user's chatting information, thereby satisfying \textit{S2}. All data between the user and the dApp, and between the dApp and Dialogflow, are transmitted over secure HTTPS channels, ensuring the confidentiality of the data. In addition, every request, except for the registration request, from the user is digitally signed with the user's private key. dApp only accepts such a request if the digital signature is successfully validated. Besides, the communications between dApp and Dialogflow are further secured with a pre-generated secret key.  All these steps combinedly fulfill \textit{S3}. Fabric, being a distributed blockchain platform, offers an effective protection against any DoS attack. A distributed network of dApps can be deployed to ensure dApp services remain available amidst a DoS attack. This satisfies \textit{S4}. We have extensively used nonces in every step of our protocol to guard against any replay attack, thereby satisfying \textit{S5}.

\vspace{1mm}
\noindent\textbf{Privacy Requirements:} Each activity related to any financial transaction, e.g., balance query or balance transfer, requires the user to sign the transaction with the private key explicitly. If the transaction is not signed, the transaction will not be considered valid and then, discarded. This implicitly represents the user consent and control for the respective transaction, thereby satisfying \textit{P1} and \textit{P2}.

\subsection{Advantages \& Limitations}
\label{sec:discussion:subsec:ad}
BONIK provides a number of advantages which are discussed next.
\begin{itemize}
    \item BONIK is the first system to integrate a chatbot with a blockchain platform enabling any user to submit financial transactions using a chatbot in a secure and privacy-friendly fashion.
    \item BONIK would be beneficial to any financial institutions in order to supplement their existing services by which their users can avail financial services. For example, instead of calling the customer care centre and being in the call centre queue for an unspecified amount of time, users could use BONIK to initiate financial transactions 24/7, any time of the day. BONIK's utility can be hugely increased by integrating with chatbot services of any social network (e.g. Facebook), thereby allowing users to access financial services from Facebook. 
    \item Being underpinned by Hyperledger Fabric means that BONIK enjoys all the essential benefits of any private blockchain platform, such as decentralisation, immutable transaction data, resiliency, transparency, automatic code execution, and so on. These features incredibly enhance the security of BONIK. The private blockchain also ensures that only authorised entities can participate in the blockchain network.
\end{itemize}

Unfortunately, the current implementation of BONIK has some limitations, as presented below:
\begin{itemize}
    \item The current PoC does not facilitate the transactions between multiple banks. However, this feature can be added by adding additional chaincode for different banks and modifying the logic of dApp and the algorithms.
    \item The current PoC utilises a small dataset to train the chatbot with only limited query language.
\end{itemize}

\subsection{Future Work}
\label{sec:discussion:subsec:future}
In future we would like to explore the following:
\begin{itemize}
    \item We would like to explore how BONIK can be integrated with Facebook chatbot service so that users can facilitate its service from Facebook.
    \item We would like to add the multiple bank feature in BONIK so that users can transact between different banks. The single channel setup deployed in the current PoC allows all nodes (peers and endorsers) of the two organisations to have full access to the blockchain data. If the multiple bank feature is integrated, this might introduce a novel privacy issue as one bank would have access to the blockchain data of another bank. This issue can be effectively addressed by connecting different banks via multiple channels of Hyperledger Fabric, thereby creating segregated transactions with separate blockchains for different banks.
    \item The 5G mobile technology is envisioned to revolutionise different service delivery models, including chatbots \cite{chatbot5g} which would provide pathways for \textit{Messaging as a Platform (MAAP)}. Within this setup, the blockchain based BONIK architecture can be the foundation upon which secure and transparent financial services can be provided. In future, we would like to pursue research in this direction as well.
\end{itemize}
\section{Conclusion}
\label{sec:conclusion}
In this paper, we have presented BONIK, a blockchain empowered chatbot for financial transactions. At first, we have formulated a set of requirements based on a rigorous threat model for financial chatbots. The architecture of BONIK has been designed to satisfy the formulated requirements and to mitigate the identified threats. We have developed a PoC prototype and described its protocol flow to show its applicability. Furthermore, we have evaluated its performance and analysed its security and privacy issues, advantages, and limitations. Using BONIK, one can execute financial transactions within a chatbot. Being rooted in a state-of-the-art private blockchain platform, Hyperledger Fabric, BONIK offers several security advantages over any existing financial chatbots. However, its true potential can be enhanced if it can be integrated with the chatbot platform in any social network, thereby laying out the foundation for a wide-scale adoption. Thus, BONIK can be regarded as a pioneering research with far-reaching potential in this domain.

\section*{Acknowledgment}
This work was in part supported by the Business Finland 5G-FORCE research project.

\begingroup
\let\itshape\upshape

\bibliographystyle{IEEEtran}
\bibliography{arxiv}

\begin{thebibliography}{10}
\providecommand{\url}[1]{#1}
\csname url@samestyle\endcsname
\providecommand{\newblock}{\relax}
\providecommand{\bibinfo}[2]{#2}
\providecommand{\BIBentrySTDinterwordspacing}{\spaceskip=0pt\relax}
\providecommand{\BIBentryALTinterwordstretchfactor}{4}
\providecommand{\BIBentryALTinterwordspacing}{\spaceskip=\fontdimen2\font plus
\BIBentryALTinterwordstretchfactor\fontdimen3\font minus
  \fontdimen4\font\relax}
\providecommand{\BIBforeignlanguage}[2]{{%
\expandafter\ifx\csname l@#1\endcsname\relax
\typeout{** WARNING: IEEEtran.bst: No hyphenation pattern has been}%
\typeout{** loaded for the language `#1'. Using the pattern for}%
\typeout{** the default language instead.}%
\else
\language=\csname l@#1\endcsname
\fi
#2}}
\providecommand{\BIBdecl}{\relax}
\BIBdecl

\bibitem{chatbot1}
\BIBentryALTinterwordspacing
R.~McGrath. (2018, May 3) ``{How} to improve customer service with chatbots''.
  Accessed: 2020-02-01. [Online]. Available:
  \url{https://chatbotsmagazine.com/ill-never-buy-from-them-again-using-chatbots-to-avoid-bad-customer-service-e6a967360244}
\BIBentrySTDinterwordspacing

\bibitem{chatbot2}
\BIBentryALTinterwordspacing
D.~Zaboj. (2020, May 6) ``{Key} chatbot statistics you should follow in 2020''.
  Accessed: 2020-07-01. [Online]. Available:
  \url{https://www.chatbot.com/blog/chatbot-statistics/}
\BIBentrySTDinterwordspacing

\bibitem{chatbotMSize}
\BIBentryALTinterwordspacing
``{Chatbot} market''. Accessed: 2020-07-01. [Online]. Available:
  \url{https://www.marketsandmarkets.com/Market-Reports/smart-advisor-market-72302363.html}
\BIBentrySTDinterwordspacing

\bibitem{okuda2018ai}
T.~Okuda and S.~Shoda, ``{AI-based} chatbot service for financial industry,''
  \emph{Fujitsu Scientific and Technical Journal}, vol.~54, no.~2, pp. 4--8,
  2018.

\bibitem{wechat}
\BIBentryALTinterwordspacing
PYMNTS. (2019, October 23) ``{WeChat} pay rolls out utility to transfer funds
  between smartphones''. Accessed: 2020-04-01. [Online]. Available:
  \url{https://www.pymnts.com/news/payment-methods/2019/wechat-pay-rolls-out-utility-to-transfer-funds-between-smartphones/}
\BIBentrySTDinterwordspacing

\bibitem{bozic2018security}
J.~Bozic and F.~Wotawa, ``Security testing for chatbots,'' in \emph{IFIP
  International Conference on Testing Software and Systems}.\hskip 1em plus
  0.5em minus 0.4em\relax Springer, 2018, pp. 33--38.

\bibitem{lai2018banking}
S.-T. Lai, F.-Y. Leu, and J.-W. Lin, ``A banking chatbot security control
  procedure for protecting user data security and privacy,'' in
  \emph{International Conference on Broadband and Wireless Computing,
  Communication and Applications}.\hskip 1em plus 0.5em minus 0.4em\relax
  Springer, 2018, pp. 561--571.

\bibitem{yan2018identifying}
F.~Yan, M.~Xu, T.~Qiao, T.~Wu, X.~Yang, N.~Zheng, and K.-K.~R. Choo,
  ``Identifying wechat red packets and fund transfers via analyzing encrypted
  network traffic,'' in \emph{TrustCom/BigDataSE 2018}.\hskip 1em plus 0.5em
  minus 0.4em\relax IEEE, 2018, pp. 1426--1432.

\bibitem{chowdhury2019comparative}
M.~J.~M. Chowdhury, M.~S. Ferdous, K.~Biswas, N.~Chowdhury, A.~Kayes,
  M.~Alazab, and P.~Watters, ``A comparative analysis of distributed ledger
  technology platforms,'' \emph{IEEE Access}, vol.~7, no.~1, pp.
  167\,930--167\,943, 2019.

\bibitem{ferdous2020blockchain}
M.~S. Ferdous, M.~J.~M. Chowdhury, M.~A. Hoque, and A.~Colman, ``Blockchain
  consensus algorithms: A survey,'' \emph{arXiv preprint arXiv:2001.07091},
  2020.

\bibitem{nakamoto2019bitcoin}
S.~Nakamoto, ``Bitcoin: A peer-to-peer electronic cash system,'' Manubot, Tech.
  Rep., 2019.

\bibitem{ferdous2019search}
M.~S. Ferdous, F.~Chowdhury, and M.~O. Alassafi, ``In search of self-sovereign
  identity leveraging blockchain technology,'' \emph{IEEE Access}, vol.~7, pp.
  103\,059--103\,079, 2019.

\bibitem{bitcoin2018}
\BIBentryALTinterwordspacing
``{Bitcoin}''. Accessed: 2020-07-10. [Online]. Available:
  \url{https://www.bitcoin.org/}
\BIBentrySTDinterwordspacing

\bibitem{ethereum2018}
\BIBentryALTinterwordspacing
``{Ethereum}''. Accessed: 2020-07-10. [Online]. Available:
  \url{https://www.ethereum.org/}
\BIBentrySTDinterwordspacing

\bibitem{litecoin2011}
\BIBentryALTinterwordspacing
``{Litecoin}''. Accessed: 2020-07-10. [Online]. Available:
  \url{https://litecoin.org/}
\BIBentrySTDinterwordspacing

\bibitem{monero2016}
\BIBentryALTinterwordspacing
``{Monero}''. Accessed: 2020-07-10. [Online]. Available:
  \url{https://www.getmonero.org/}
\BIBentrySTDinterwordspacing

\bibitem{hyperledger2018}
\BIBentryALTinterwordspacing
``{Hyperledger}''. Accessed: 2020-07-10. [Online]. Available:
  \url{https://www.hyperledger.org/}
\BIBentrySTDinterwordspacing

\bibitem{quorum2018}
\BIBentryALTinterwordspacing
``{Quorum Blockchain}''. Accessed: 2020-07-10. [Online]. Available:
  \url{https://www.goquorum.com/}
\BIBentrySTDinterwordspacing

\bibitem{abdul2015survey}
S.~A. Abdul-Kader and J.~Woods, ``Survey on chatbot design techniques in speech
  conversation systems,'' \emph{International Journal of Advanced Computer
  Science and Applications}, vol.~6, no.~7, 2015.

\bibitem{siri2020}
\BIBentryALTinterwordspacing
``{Apple Siri}''. Accessed: 2020-08-02. [Online]. Available:
  \url{https://www.apple.com/siri/}
\BIBentrySTDinterwordspacing

\bibitem{roman}
\BIBentryALTinterwordspacing
``{Roman saves her mom}''. Accessed: 2020-02-02. [Online]. Available:
  \url{https://www.cnet.com/news/child-saves-mother-iphone-siri-uk/}
\BIBentrySTDinterwordspacing

\bibitem{googleAssistant2020}
\BIBentryALTinterwordspacing
``{Google Assistant}''. Accessed: 2020-08-02. [Online]. Available:
  \url{https://assistant.google.com/}
\BIBentrySTDinterwordspacing

\bibitem{alexa2020}
\BIBentryALTinterwordspacing
``{Amazon Alexa}''. Accessed: 2020-08-02. [Online]. Available:
  \url{https://alexa.amazon.com/}
\BIBentrySTDinterwordspacing

\bibitem{classification}
\BIBentryALTinterwordspacing
J.~Grills. (2019, May 15) ``{Is} voice activated chatbot better than the
  text-based chatbot?''. Accessed: 2020-06-01. [Online]. Available:
  \url{https://chatbotsmagazine.com/is-voice-activated-chatbot-better-than-the-text-based-chatbot-7230e9161620}
\BIBentrySTDinterwordspacing

\bibitem{brill1995transformation}
E.~Brill, ``Transformation-based error-driven learning and natural language
  processing: A case study in part-of-speech tagging,'' \emph{Computational
  linguistics}, vol.~21, no.~4, pp. 543--565, 1995.

\bibitem{bakshi2016opinion}
R.~K. Bakshi, N.~Kaur, R.~Kaur, and G.~Kaur, ``Opinion mining and sentiment
  analysis,'' in \emph{3rd INDIACom}.\hskip 1em plus 0.5em minus 0.4em\relax
  IEEE, 2016, pp. 452--455.

\bibitem{finChatbot}
\BIBentryALTinterwordspacing
J.~Tarbal. (2020, January 27) ``{Chatbots} in financial services: Benefits, use
  cases and key features''. Accessed: 2020-06-01. [Online]. Available:
  \url{https://www.artificial-solutions.com/blog/chatbots-financial-services-benefits-use-cases}
\BIBentrySTDinterwordspacing

\bibitem{shostack2014threat}
A.~Shostack, \emph{Threat modeling: Designing for security}.\hskip 1em plus
  0.5em minus 0.4em\relax John Wiley \& Sons, 2014.

\bibitem{HyperledgerFabric}
\BIBentryALTinterwordspacing
``{Hyperledger Fabric}''. Accessed: 2020-07-10. [Online]. Available:
  \url{https://www.hyperledger.org/use/fabric}
\BIBentrySTDinterwordspacing

\bibitem{nodejs}
\BIBentryALTinterwordspacing
``{Node.js}''. Accessed: 2020-07-10. [Online]. Available:
  \url{https://nodejs.org/en/}
\BIBentrySTDinterwordspacing

\bibitem{express}
\BIBentryALTinterwordspacing
``{Express JS}''. Accessed: 2020-07-10. [Online]. Available:
  \url{https://expressjs.com/}
\BIBentrySTDinterwordspacing

\bibitem{dialogflow}
\BIBentryALTinterwordspacing
``{Google Dialogflow}''. Accessed: 2020-07-10. [Online]. Available:
  \url{https://cloud.google.com/dialogflow}
\BIBentrySTDinterwordspacing

\bibitem{caliper}
\BIBentryALTinterwordspacing
``{Hyperledger Caliper}''. Accessed: 2020-07-10. [Online]. Available:
  \url{https://www.hyperledger.org/use/caliper}
\BIBentrySTDinterwordspacing

\bibitem{chatbot5g}
\BIBentryALTinterwordspacing
C.~Knight. (2020, May 8) ``{Chatbots} getting the {5G} treatment as networks go
  live''. Accessed: 2020-06-01. [Online]. Available:
  \url{https://thechatbot.net/chatbots-5g/}
\BIBentrySTDinterwordspacing

\end{thebibliography}
\endgroup
\end{document}